\documentclass[aps,prd,amsfonts,preprint,
eqsecnum,nofootinbib]
{revtex4}
\usepackage{graphics,epsfig}
\begin{document}
\preprint{WM-02-103}
%
\title{\vspace*{0.3in}
Bulk Majorons at Colliders
\vskip 0.1in}
\author{Christopher D. Carone}
\email[]{carone@physics.wm.edu}
\author{Justin M. Conroy}
\email[]{jconroy@camelot.physics.wm.edu}
\author{Herry J. Kwee}
\email[]{herry@camelot.physics.wm.edu}

\affiliation{Nuclear and Particle Theory Group, Department of
Physics, College of William and Mary, Williamsburg, VA 23187-8795}
\date{April, 2002}
\begin{abstract}
Lepton number violation may arise via the spontaneous breakdown of a global 
symmetry.  In extra dimensions, spontaneous lepton number violation in the bulk 
implies the existence of a Goldstone boson, the majoron $J^{(0)}$, as well as 
an accompanying tower of Kaluza-Klein (KK) excitations, $J^{(n)}$.  Even if the 
zero-mode majoron is very weakly interacting, so that detection in low-energy processes 
is difficult, the sum over the tower of KK modes may partially compensate in
processes of relevance at high-energy colliders.  Here we consider the inclusive 
differential and total cross sections for $e^-e^- \rightarrow W^- W^- J$, where $J$ 
represents a sum over KK modes.  We show that allowed parameter choices exist
for which this process may be accessible to a TeV-scale electron collider.
\end{abstract}
\maketitle


\section{Introduction}\label{sec:intro}

Perhaps the simplest mechanism of lepton number violation is the spontaneous
breakdown of a global symmetry.  This possibility may be of relevance in nature
if neutrino masses are lepton number violating.  The associated Goldstone boson,
the majoron, has phenomenological consequences that have been studied in the context
of a variety of models~\cite{models,mm,mmm}.  The majoron can be produced in neutrinoless
double beta decay~\cite{ggn}, leading to a distinctive form for the electron sum-energy 
spectrum. In addition, majorons might be discerned through their astrophysical effects, in 
particular, by altering the total neutrino luminosity from supernovae, and possibly 
also the energy spectra of different supernova neutrino flavors~\cite{tomas}.   

Most majoron models have been formulated in the context of purely four-dimensional field
theories.  Recent interest in the possibility of extra compactified spatial dimensions~\cite{led}
has led to new variations~\cite{mohapatra,triplets} on the conventional majoron scenarios.  
Mohapatra, {\em et al.,} \cite{mohapatra} have noted that a bulk majoron field in a 
five-dimensional theory leads to additional contributions to neutrinoless double beta decay 
from the majoron Kaluza-Klein (KK) excitations that are kinematically accessible to the final 
state.  Such additional contributions would enhance the decay rate in a singlet majoron model
so that effects could be manifest at a next-generation double beta decay experiment.  
In such a scenario, the additional KK contributions to the decay alter the shape
of the electron sum energy spectrum ({\em i.e.} the spectral index) in a way that
would distinguish it from that of a conventional majoron model.

In this note, we consider another possible signal for a singlet majoron that is defined on
a compactified, extra-dimensional space.  We will first assume that the scale of 
compactification $1/R\equiv \mu_0$ is such that the lightest KK excitation is too heavy to 
participate in double beta decay, so that the effects discussed in 
Ref.~\cite{mohapatra} are absent. In addition, the compactification scale will be 
high enough in comparison to typical supernova core temperatures ($\mu_0 > 30$~MeV) to 
evade additional astrophysical bounds on the KK modes. In other words, we restrict 
ourselves to a choice of parameters in which the lightest KK modes are not directly 
constrained by the same low-energy and astrophysical bounds as the zero mode.  We 
then focus on high-energy processes in which the sum over majoron KK modes may still 
lead to a substantial enhancement.  Specifically, we consider the process 
$e^- e^- \rightarrow W^- W^- J$, where $J$ represents a sum over undetected modes.
Searches for neutrinoless double beta decay require that the $J^{(0)}$ coupling to 
neutrinos is extremely small ($< O(10^{-5})$)~\cite{tomas}, so that the production rate 
for an individual mode $J^{(n)}$ is completely negligible.  Nevertheless, the number 
of KK modes between  $\sim 30$~MeV and the TeV scale can be 
enormous and provides a partially compensating effect.  Our goal is to study this issue 
quantitatively in a simplified one generation majoron model in which the majoron can
propagate in $\delta$ extra, compactified spatial dimensions.  We will see that in the
case where $\delta=1$, allowed parameter choices exist where $O(10^2)$ events per year
might be produced at an NLC, while for larger $\delta$ the result is significantly 
suppressed. 

\begin{figure}[ht]
\centerline {\epsfxsize 4. in \epsfbox{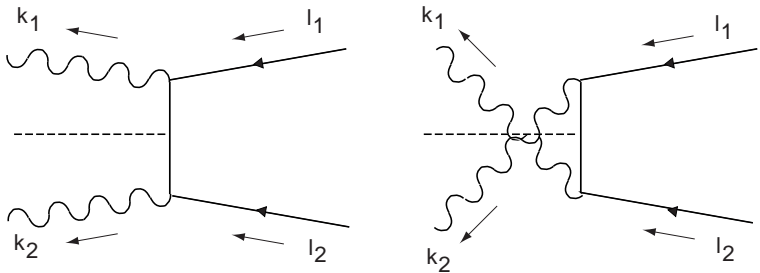}   }
\caption{Feynman diagrams for $e^- e^- \rightarrow W^- W^- J$}
\label{feyndiags}
\end{figure}
\section{Effective Four Dimensional Lagrangian} \label{sec:two}

We consider a singlet majoron model, in which lepton number is a global symmetry that
is broken spontaneously by a gauge singlet Higgs field $\chi$. The singlet is 
assigned $-2$ units of lepton number; operators of the form 
$\chi (LH)^2$, where $L$ is a weak doublet lepton field and $H$ the standard model 
Higgs, are allowed in the high energy theory and generate neutrino majorana 
masses when both $\chi$ and $H$ obtain vacuum expectation values (vevs).

We study an extra-dimensional extension of the model in which the majoron may propagate 
in the bulk, while all other standard model fields are confined to an orbifold fixed 
point.  (We describe the orbifold below.)  In this context, the operator of interest
may be written explicitly as
\begin {equation}
{\cal L}^{4D} (x) = \int d^{\delta} y \, \frac {c}{M^{2+
\delta/2}} \chi {\overline{L^c}}^{\alpha}H^\beta  L^\rho H^\eta \epsilon_{\alpha\beta}
\epsilon_{\rho\eta}\, \delta^{\delta}(y) ,
\label{eq:theop}
\end {equation}
where $\alpha, \beta, \rho,$ and $\eta$ are weak SU(2) indices, and the superscript $c$
indicates charge conjugation, $\psi^c \equiv C \, {\bar\psi}^T$, where $C$ is the
charge conjugation matrix. The coupling $c$ is undetermined in the effective theory 
and $M$ represents the scale of new physics, which we identify as the reduced Planck
scale for the theory.  As written in Eq.~(\ref{eq:theop}), the field $\chi$ 
has mass dimension $1+\delta/2$, where $\delta$ is the number of extra dimensions; all
other fields have their usual four-dimensional normalization.

We first take into account spontaneous symmetry breaking and then integrate over the
extra dimensions.  One may assume simple potentials to assure that both $H$ and
$\chi$ develop vacuum expectation values.  For example, with 
\begin {equation}
V(\chi) = -\frac {{m_\chi}^2}{2} \chi^\dagger \chi + \frac{\lambda}{4M^\delta} 
\left( \chi^\dagger \chi\right)^2
\end {equation}
one finds that
\begin {equation}
\left<\chi\right> \equiv v_\chi = \frac {m_\chi M^{\delta/2}}{\sqrt{\lambda}} .
\end {equation}
We choose instead a model-independent approach and simply assume that suitable 
symmetry breaking sectors exist.  We thus set
\begin{equation}
H = \frac{1}{\sqrt{2}} \left( \begin{array} {c} 0 \\ {v} \end{array} 
\right) 
\end{equation}
where $v\approx 250$~GeV, and expand
\begin{equation}
\chi = \frac {1}{\sqrt{2}} \left( v_\chi + \varphi + iJ \right)  \,\,\, .
\end{equation}
Henceforth we focus on the pseudoscalar $J$ whose zero mode component corresponds
to the conventional massless majoron field.

From Eq.~(\ref{eq:theop}) it now follows that a neutrino majorana mass term is 
generated,
\begin {equation}
{\cal L}^{4D} = \frac{1}{2} m_\nu \overline{\nu^c} \, \nu
\end{equation}
where
\begin{equation}
m_\nu = \frac{c \, v^2}{M^{2+\delta/2}} \frac{ v_\chi}{\sqrt{2}} \,\,\, .
\label{eq:numass}
\end{equation}
Similarly, the majoron coupling 
\begin {equation}
{\cal L}^{4D} = \int d^{\delta} y \frac{i}{2\sqrt{2}}
\frac{c v^2}{M^{2+\delta/2}}  J \overline{\nu^c} \gamma^5 \nu \, \delta^\delta(y)
\label{eq:mcoup}
\end {equation}
can now be simplified by substituting the KK decomposition of the field J. For
compactification on a circle, for example, we would expand
\begin {equation}
J (x^\mu,\vec{y}) = \sum _{n_1=-\infty}^{\infty} 
\sum _{n_2=-\infty}^{\infty} \cdots \sum _{n_\delta=-\infty}^{\infty}
J^{(\vec{n})} (x^\mu)\, \exp \left(\frac{i \, \vec{n} \cdot \vec{y}}{R}\right)
\end {equation}
where $\vec{n}=(n_1,n_2,\ldots n_\delta)$ and $R$ is the radius
of compactification.  Since we would like to confine all fields other than
$J$ to a 4D subspace we compactify instead on the orbifold $S^\delta/ Z_2$,
which effectively cuts the extra-dimensional space in half.  Even and odd modes
under this $Z_2$ may be written
\begin{equation}
J_+ (x^\mu, \vec{y}) = \sum_{\vec{n} \ge \vec{0}} J_+^{(n)}(x^\mu) \cos \left(
\frac{ \, \vec{n}\cdot\vec{y}}{R}\right) 
\label{eq:even}
\end{equation}
\begin{equation}
J_- (x^\mu, \vec{y}) = \sum_{\vec{n}> \vec{0}} J_-^{(n)}(x^\mu) \sin \left(
\frac{ \, \vec{n}\cdot\vec{y}}{R} \right)
\label{eq:odd}
\end{equation}
The vector $\vec{n}$ may either be identically $\vec{0}$ 
or it may be ``positive", $\vec{n}>\vec{0}$:  writing the vector as a row, a positive 
vector is one whose first nonzero element reading from left to right is positive while
all remaining elements have values that are unrestricted~\cite{ncpaper}.  The sum in
Eq.~(\ref{eq:odd}) is over positive $\vec{n}$, excluding the zero mode. 

We assume that the majoron field is even under the orbifold parity so that couplings
to matter at the fixed point $\vec{y}=\vec{0}$ are non-vanishing.  In addition to
substituting Eq.~(\ref{eq:even}) into Eq.~(\ref{eq:mcoup}) and integrating over
$\vec{y}$, we also must rescale the majoron field to obtain canonically normalized
4D kinetic terms $J_+^{(\vec{n})} 
\rightarrow \zeta J_+^{(\vec{n})}/(2 \pi R)^{\delta/2}$  where $\zeta=1$ for the zero mode 
and $\zeta=\sqrt{2}$ otherwise. Henceforth, $J_+$ will refer to the conventionally
normalized 4D field.  We finally obtain
\begin {equation}
{\cal L}^{4D} = \frac {i}{2} \xi \, \overline{\Psi} \gamma^5 \Psi
\left[J^{(0)}_+ + \sqrt{2} \sum _{\vec(n)} J^{(n)}_+
\right] \,\,\, ,
\end {equation}
where
\begin{equation}
\xi = \frac{\Lambda^{\delta/2} m_\nu}{v_\chi} \,\,\, .
\label{eq:xidef}
\end{equation}
Here $\Lambda = 1/(2 \pi R)$ and $\Psi$ is a four-component majorona spinor 
satisfying $\Psi^c=\Psi$ with $\Psi_L \equiv \nu_L$.  Note that $\xi$ is
dimensionless since $v_\chi$ has mass dimension $1+\delta/2$.

We let phenomenological considerations dictate our choice of model parameters.  The 
coupling $\xi/2 \equiv g$ is nothing more that the conventional zero-mode majoron 
coupling to neutrinos.  Constraints on neutrinoless double beta decay restrict 
$g<3 \times 10^{-5}$ while those on supernova majoron luminosity eliminate an interval
$3 \times 10^{-7} < g < 2 \times 10^{-5}$~\cite{tomas}.  For the cases of $\delta=1$
or $2$ extra dimensions, we will avoid these excluded regions for a mildly large
value of the operator coupling, $c = 4$.  In addition, we take the compactification 
scale $1/R=100$~MeV which is larger that the typical end point energies in neutrinoless
double beta decay ($\sim 3$~MeV) and supernovae core temperatures ($\sim 30$~MeV).  This
assures that there are no further restrictions on $\xi$ from the KK majoron states.
Since we will be primarily interested in effects at a $\sqrt{s}=1.5$~TeV NLC, we will
take the cutoff of our theory to be somewhat higher, $M=3$~TeV\footnote{We assume only
that the majoron lives in $\delta$ extra dimensions of radii $\mu_0^{-1}$; this does not
preclude other larger extra dimensions accessible to gravity only.  Hence the value of
$M$ is not fixed by the other parameter choices in our calculation.}.  
By combining  Eqs.~(\ref{eq:numass}) and (\ref{eq:xidef}) we then 
find $\xi=4.6 \times 10^{-5}$ and $\xi=1.0 \times 10^{-7}$ for $\delta=1$ and $2$, 
respectively.  In either case, desired values of $m_\nu$ may be obtained by adjusting
$v_\chi$; we take $m_\nu=0.1$~eV in our estimates below.

\section{Cross Sections} \label{sec:three}

    To calculate the cross section for $e^{-}e^{-}\rightarrow W^{-}W^{-}J$, we 
evaluate the diagrams shown in Fig.~\ref{feyndiags}.  Self-conjugacy of the majorana 
neutrino field $\Psi$ leads to three distinct forms for the neutrino propagators.
Summing over these, the majoron-neutrino vertex and internal lines
in these diagrams take on the simplified form
\begin{equation}
\frac{\xi}{(p_1^{2}-m_\nu^{2})(p_2^{2}-m_\nu^{2})}{(\not\!p_1^{T}+m_\nu) \gamma^5
(i \, \gamma^{0}\gamma^{2})(\not\!p_2+m_\nu)}
\end{equation}
where $p_1$ and $p_2$ are momenta identified with $l_1-k_1$ and $l_2-k_2$, respectively, 
in the first diagram of Fig.~\ref{feyndiags}.  Using this result, one may calculate
the spin-averaged invariant amplitude $|M^2|=|M_1+M_2|^2$:
\[\mid M_{1} \mid^{2}=\frac{\zeta^{2}\xi^{2}g_{w}^{4}}{16(p^{2}-m_\nu^{2})^{2}
(q^{2}-m_\nu^{2})^{2}}
(\frac{k_{1}^{\alpha}k_{1}^{\mu}}{m_{w}^2}-g^{\alpha\mu})(\frac{k_{2}^{\beta}k_{2}^{\nu}}
{m_{w}^2}-g^{\beta\nu})\]
\begin{equation}
\times Tr{\left\{\not\!l_{1} \gamma^{\mu} \! \not\!p \not\! q \, \gamma^{\nu} 
\!  \not\!l_{2} \,
\gamma^{\beta}\!\not\!q\not\!p \,(\frac{1-\gamma^{5}}{2})\gamma^{\alpha}\right\} },
\label{m1}
\end{equation}
{\samepage
\[ \mid M_{2} \mid^{2}=\frac{\zeta^{2}\xi^{2}g_{w}^{4}}{16(\tilde{p}^{2}-m_\nu^{2})^{2}
(\tilde{q}^{2}-m_\nu^{2})^{2}}(\frac{k_{1}^{\alpha}k_{1}^{\mu}}{m_{w}^2}-g^{\alpha\mu})
(\frac{k_{2}^{\beta}k_{2}^{\nu}}{m_{w}^2}-g^{\beta\nu})\]
\begin{equation}    
\times Tr{\left\{ \not\!l_{1} \gamma^{\nu} \! \not\!\tilde{p}\not\!\tilde{q}\,
\gamma^{\mu}\!\not\!l_{2} \gamma^{\alpha}\! \not\!\tilde{q}\not\!\tilde{p}\,
(\frac{1-\gamma^{5}}{2})\gamma^{\beta}\right\}},
\label{m2}
\end{equation}
}
\[\mid M_{1} M_{2}^{\ast} \mid=\frac{\zeta^{2}\xi^{2}g_{w}^{4}}{16(p^{2}-m_\nu^{2})
(q^{2}-m_\nu^{2})
(\tilde{p}^{2}-m_\nu^{2})(\tilde{q}^{2}-m_\nu^{2})}(\frac{k_{1}^{\alpha}k_{1}^{\mu}}
{m_{w}^2}-g^{\alpha\mu})(\frac{k_{2}^{\beta}k_{2}^{\nu}}{m_{w}^2}-g^{\beta\nu})\]
\begin{equation}
\times Tr{\left\{\not\!l_{1} \gamma^{\mu}\! \not\!p\not\!q\,\gamma^{\nu}\!\not\!l_{2} 
\gamma^{\alpha}\!\not\!\tilde{q}\not\!\tilde{p}\,(\frac{1-\gamma^{5}}{2})\gamma^{\beta}
\right\}} \,\,\, ,
\label{m1m2}
\end{equation}
where $p=l_{1}-k_{1}$, $q=l_{2}-k_{2}$, $\tilde{p}=l_{1}-k_{2}$, and 
$\tilde{q}=l_{2}-k_{1}$.
In Eqs.~(\ref{m1})-(\ref{m1m2}), we have averaged over the electron spins and 
summed over the $W^{-}$ polarizations.  The explicit form of $|M|^2$ was computed
symbolically using FeynCalc~\cite{fc} and is somewhat cumbersome, so we only present 
numerical results below.  The differential cross section for producing one majoron mode of 
mass $m_{J}$ may be written
\begin{equation}
\frac{d\sigma_J}{dE_{1} dE_{2} d(\cos\theta_{1}) d(\cos\theta_{2})}=
\frac{1}{256 \pi^{4} \, s}\frac{1}{\sin\theta_{1}\sin\theta_{2}\sin\phi} | M |^{2}
\label{xsect}
\end{equation}
where $\sqrt{s}$, $E_{1}$, and $E_{2}$ are the center of mass and $W^{-}$ energies, 
respectively.  The angles $\theta_i$ are defined relative to the beam direction,
while the angle $\phi$ is the azimuthal separation between the two $W$'s.  
Note that $\cos\phi$ is fixed by the energy-conserving delta function in the 
Lorentz-invariant phase space,
\begin{equation}
\cos\phi= \frac{(\sqrt{s}-E_1-E_2)^2-m_J^2-p_1^2-p_2^2-2 p_1 p_2 \cos\theta_1\cos
\theta_2}{2 p_1 p_2 \sin\theta_1\sin\theta_2}  \,\,\, ,
\label{eq:cosphi}
\end{equation}
where $p_i$ are the $W$ momenta and $m_J$ is the mass of a given majoron KK mode.

The process of interest to us, however, is not the production of a single KK mode,
but a sum over all KK modes that are kinematically accessible.  We therefore
multiply Eq.~(\ref{xsect}) by the multiplicity of KK modes per unit mass.  Hence,  
\begin{equation}
\frac{d\sigma}{dE_1 dE_2 d(\cos\theta_{1}) d(\cos\theta_{2}) dm_j}= 
\frac{1}{2}\frac{2\pi^{\frac{\delta}{2}}m_j^{\delta-1}}
{\mu_0^{\delta}\Gamma(\frac{\delta}{2})}\, 
\frac{d\sigma_J}{dE_{1} dE_{2} d(\cos\theta_{1}) d(\cos\theta_{2})}
\label{eq:fform}
\end{equation}
The new multiplicative factor can be derived from the surface area of 
a $\delta$-dimensional sphere in $\mu_0 \vec{n}$ space, where $m_j = \mu_0 |\vec{n}|$.
The additional factor of $1/2$ corrects for overcounting arising from the 
restriction on the values of $\vec{n}$ in Eq.~(\ref{eq:even}).
\begin{figure}[t]
\centerline {\epsfxsize 4. in \epsfbox{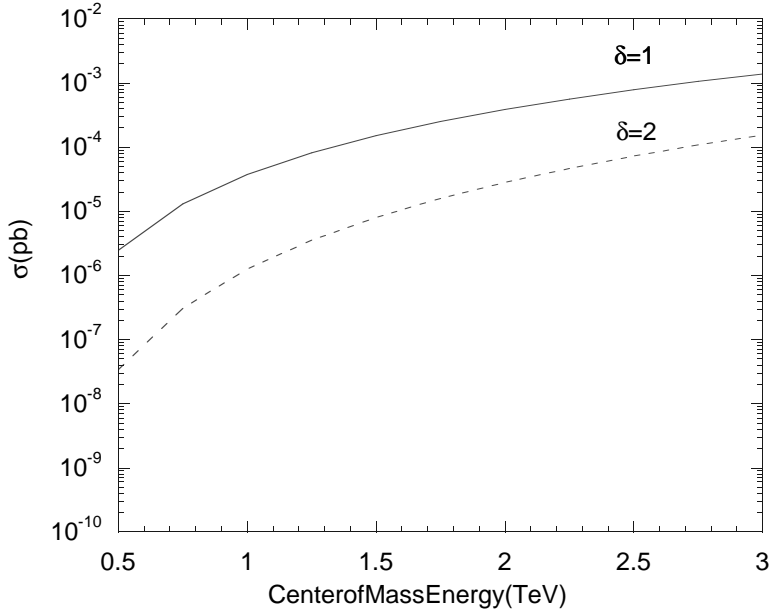}   }
\caption{Total cross section for $e^- e^- \rightarrow W^- W^- J$}
\label{fig:xsec}
\end{figure}

\begin{figure}[t]
\centerline {\epsfxsize 4. in \epsfbox{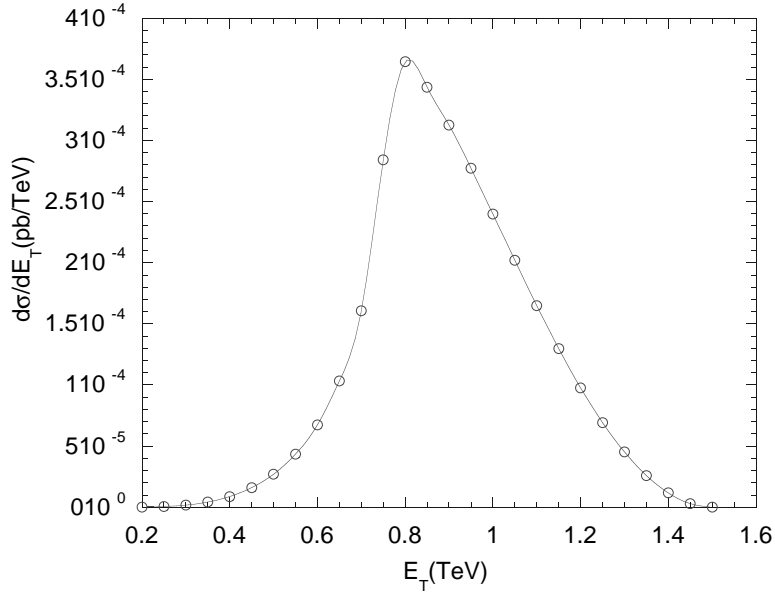}   }
\caption{$E_T$ distribution for $e^- e^- \rightarrow W^- W^- J$, for $\delta=1$
and $\sqrt{s}=1.5$~TeV}
\label{fig:etdist}
\end{figure}

Total and differential cross sections may be obtained by integrating Eq.~(\ref{eq:fform}).
We evaluated the total cross section $\sigma(\sqrt{s})$ and the $W$ sum energy 
distribution $d\sigma/dE_T$, where $E_T \equiv E_1 + E_2$, for the parameter choices
described at the end of Section~\ref{sec:two}. Numerical integration
was performed using three independent Monte Carlo integration packages (Mathematica, 
the CERNLIB routine DIVON4~\cite{d4}, and the numerical recipes package VEGAS~\cite{nr}), 
and consistent results were obtained.  Fig.~\ref{fig:xsec} shows the total cross section 
in picobarns as a function of center of mass energy for one and two extra dimensions.  A 
cut on $\cos\theta_i < 0.8$ was employed to eliminate events where one or more of the $W$'s
are close to the beam direction.  We do not find that our results vary strongly with this 
choice.  Conveniently, this cut also eliminates a kinematical region 
where there are infrared divergences that must be cancelled by including higher-order 
corrections.  For a $1.5$~TeV NLC with a luminosity of  
$3.4 \times 10^{34}$~cm$^{-2}$~s$^{-1}$~\cite{nlc}, we find that one year of running yields
$164$ and $9$ events for one and two extra dimensions, respectively.  For larger
values of $\delta$ the results are further suppressed so that detection at a TeV scale
machine looks less promising.  In the $\delta=1$ case, we display the $W$ sum energy
spectrum in Fig.~\ref{fig:etdist}, which may be useful in distinguishing majoron 
production from other physics backgrounds, in particular, from those in which missing
energy is not expected.

\section{Discussion}

In this letter we have considered possible phenomenological consequences of a bulk majoron
scenario, in a limit where the Kaluza-Klein excitations of the majoron $J^{(n)}$ are too
heavy to have an effect on double beta decay measurements. Exploiting the analogy to 
graviton production, we entertained the possibility that the sum
over KK modes would be sufficient to compensate for the weakness of the majoron-neutrino
coupling and lead to potentially observable collider signatures.  We showed the the rate 
for the process $e^- e^- \rightarrow W^- W^- J$, where $J$ represents the sum over majoron 
KK modes that go undetected experimentally, is not negligible at a TeV-scale lepton 
collider.  In addition, we presented numerical results for the $W$ sum-energy spectrum,
which might be relevant experimentally.

It is worth pointing out that one possible physics background to the majoron production 
process we consider is $e^- e^- \rightarrow W^- W^-$, which is lepton number violating 
and does not occur in the standard model. However, this background may arise due to other 
nonstandard sources of lepton number violation in the theory: the presence of an extra, 
heavy majorona neutrino, for example, can lead to $W^- W^-$ production via its $t$-channel 
exchange~\cite{london}, while the $s$-channel exchange of an exotic doubly-charged Higgs 
in some models~\cite{dch} can have  the same effect. Distinguishing KK majoron production 
from these possibilities should be easy since majoron production implies a nontrivial $W^-$ sum 
energy spectrum, while the other cases yield no missing energy if the $W$'s are completely
reconstructed.  Standard model backgrounds like $e^- e^- \rightarrow e^- e^- W^+ W^-$ generally 
require that both final state electrons are missed {\em and} that the $W$ charges are not 
identified.  If both $W$ bosons decay leptonically, adequate charged lepton identification is 
required to suppress this background in determining total event rates.  To find angular or
sum energy distributions, it is instead desirable to consider the decays of both $W$ bosons to 
jets, so that the $W$ momenta can be completely reconstructed.  In this case it may still be 
possible to reduce the background from missed final-state electrons by using jet charge 
identification techniques to determine the $W$ charges.  While this is nontrivial, the use of 
neural networks in extracting the charges in $W$ pair production appears to be a promising 
approach~\cite{mostafa}.

Finally, we note that the discussion in this letter has focused on a one-generation model
while realistic three-generation theories would necessarily involve lepton-flavor changing
effects.  A phenomenological analysis like that in Ref.~\cite{tomas}, that takes into
account simultaneous bounds on the majoron coupling and on the neutrino mass matrix, would 
be well motivated.  Explicit model building may also be useful in explaining the choice of 
parameters and scales that were selected in the present analysis for their 
phenomenological relevance, and would provide a framework for addressing cosmological issues.  
In any case, our immediate results suggests that the future collider searches for lepton number 
violation in the ``inverse beta decay"  channel may have a renewed, extra-dimensional motivation.

%
\begin{acknowledgments}
We thank the National Science Foundation for support  under Grant No.\ PHY-9900657 
and the Jeffress Memorial Trust for support under Grant No.~J-532.
\end{acknowledgments}


\begin{thebibliography}{99}

\bibitem{models}
Y.~Chikashige, R.~N.~Mohapatra and R.~D.~Peccei,
Phys.\ Lett.\ B {\bf 98}, 265 (1981);
G.~B.~Gelmini and M.~Roncadelli,
Phys.\ Lett.\ B {\bf 99}, 411 (1981).

\bibitem{mm}
A.~Masiero and J.~W.~Valle,
Phys.\ Lett.\ B {\bf 251}, 273 (1990).

\bibitem{mmm}
C.~P.~Burgess and J.~M.~Cline,
Phys.\ Rev.\ D {\bf 49}, 5925 (1994)
[arXiv:hep-ph/9307316];
C.~P.~Burgess and J.~M.~Cline,
Phys.\ Lett.\ B {\bf 298}, 141 (1993)
[arXiv:hep-ph/9209299];
C.~D.~Carone,
Phys.\ Lett.\ B {\bf 308}, 85 (1993)
[arXiv:hep-ph/9302290].

\bibitem{ggn}
H.~M.~Georgi, S.~L.~Glashow and S.~Nussinov,
Nucl.\ Phys.\ B {\bf 193}, 297 (1981).

\bibitem{tomas}
R.~Tomas, H.~Pas and J.~W.~Valle,
Phys.\ Rev.\ D {\bf 64}, 095005 (2001)
[arXiv:hep-ph/0103017];
M.~Kachelriess, R.~Tomas and J.~W.~Valle,
Phys.\ Rev.\ D {\bf 62}, 023004 (2000)
[arXiv:hep-ph/0001039].

\bibitem{led}
I.~Antoniadis,
Phys.\ Lett.\ B {\bf 246}, 377 (1990);
J.~D.~Lykken,
Phys.\ Rev.\ D {\bf 54}, 3693 (1996);
N.~Arkani-Hamed, S.~Dimopoulos and G.~R.~Dvali,
Phys.\ Lett.\ B {\bf 429}, 263 (1998);
Phys.\ Rev.\ D {\bf 59}, 086004 (1999);
I.~Antoniadis, N.~Arkani-Hamed, S.~Dimopoulos and G.~R.~Dvali,
Phys.\ Lett.\ B {\bf 436}, 257 (1998);
K.~R.~Dienes, E.~Dudas and T.~Gherghetta,
Nucl.\ Phys.\ B {\bf 537}, 47 (1999);
K.~R.~Dienes, E.~Dudas and T.~Gherghetta,
Phys.\ Lett.\ B {\bf 436}, 55 (1998).

\bibitem{mohapatra}
R.N. Mohapatra, A. P\'{e}rez-Lorenzana, and C.A. de S. Pires,
Phys.\ Lett.\ {\bf B491} (2000) 143.

\bibitem{triplets}
E.~Ma, M.~Raidal and U.~Sarkar,
Nucl.\ Phys.\ B {\bf 615}, 313 (2001) [arXiv:hep-ph/0012101];
Phys.\ Rev.\ Lett.\  {\bf 85}, 3769 (2000)
[arXiv:hep-ph/0006046].

\bibitem{ncpaper}
See, for example, the discussion in C.~E.~Carlson and C.~D.~Carone,
Phys.\ Rev.\ D {\bf 65}, 075007 (2002)
[arXiv:hep-ph/0112143].

\bibitem{fc}
R.~Mertig, M.~Bohm and A.~Denner,
Comput.\ Phys.\ Commun.\  {\bf 64}, 345 (1991).

\bibitem{d4} 
J.~H.~Friedman and M.~H.~Wright,
``Divonne4: A Program For Multiple Integration And Adaptive Importance Sampling,''
CGTM-193-REV.

\bibitem{nr}
B.P. Flannery, W.H. Press, S.A. Teukolsky, ``Numerical Recipes in Fortran 77'',
2nd edition, Cambridge University Press, Cambridge 1992.

\bibitem{nlc}
[NLC Collaboration],
in {\it Proc. of the APS/DPF/DPB Summer Study on the Future of Particle Physics 
(Snowmass 2001) } ed. R.~Davidson and C.~Quigg, SLAC-R-571.

\bibitem{london}
T.~Rizzo, Phys.\ Lett.\ B {\bf 116}, 23 (1982);
D.~London, G.~Belanger and J.~N.~Ng,
Phys.\ Lett.\ B {\bf 188}, 155 (1987);
D.~A.~Dicus, D.~D.~Karatas and P.~Roy,
Phys.\ Rev.\ D {\bf 44}, 2033 (1991);
C.~A.~Heusch and P.~Minkowski,
Nucl.\ Phys.\ B {\bf 416}, 3 (1994).

\bibitem{dch}
T.~G.~Rizzo,
Phys.\ Rev.\ D {\bf 25}, 1355 (1982)
[Addendum-ibid.\ D {\bf 27}, 657 (1982)].

\bibitem{mostafa}
M. Mjahed, Nucl.\ Inst.\ Meth.\ {\bf A449} (2000) 602.





\end{thebibliography}

\end{document}